# Delivery of Liquid Metal to the Target Vessels as Vascular Embolic Agent to Starve Diseased Tissues or Tumors to Death

Qian Wang [1,3], Yang Yu [1,3] and Jing Liu [1,2]*

1. Department of Biomedical Engineering, School of Medicine,
Tsinghua University, Beijing 100084, China.
2. Technical Institute of Physics and Chemistry, Chinese Academy of Sciences,
Beijing 100190, China.
3. These authors contributed equally to this work.

Correspondence should be addressed to J.L. (jliubme@tsinghua.edu.cn)

**Abstract**

Tumor growth strongly depends on the continuous blood and nutrients supply. Theoretically, it is an ideal therapeutic way of treating tumor by vascular embolization, which also reduces the harmful effects on surrounding normal tissues. However, most of the currently available vascular embolic agents are still rather insufficient in performances to fulfill the real clinical need due to the reasons like: incomplete filling of target vasculature, being easily washed away by blood or body solution, or just producing toxicity to human body. Here from an alternative way, the body temperature liquid metal, a kind of soft and highly compliant material, was proposed for the first time as a new conceptual blood vessel embolization strategy towards tumor physical therapy. With the unique capability of easy phase transition between liquid and solid state and sub-cooling behavior, such agent can be fluently injected into and to fully fill the tiny vessels including the ending capillaries. The *in vitro* cytotoxicity experiments have been performed which indicated that treating localized diseased tissues through liquid metal embolic agent is acceptable. Endowed with a high density, the liquid metal-filled vessels are highly visible under the CT scan, which offers the potential of the diagnosis-treatment integration with only one single material. To further demonstrate the new liquid metal vascular embolization therapy, several typical experiments *in vivo* on vasculatures of rabbit ears and mouse tails have been performed to provide the evidences of destroying the targeted tissues due to liquid metal filling. To interpret the liquid metal starvation therapy effects, theoretical model has been established to simulate the tumor growth with zero, partial or complete filling of the metal agent inside the vessels. All the results support that, given appropriate administration, the liquid metal embolization is able to destruct the target regions effectively and might be able to starve the tumors to death through a relatively easy and compliant way. This study lays the foundation of a promising tumor starvation therapy with both high contrast image guidance capability and overall treatment outcome.

**Keywords:** Liquid metal; Tumor treatment; Blood vessel; Embolic agent; Starvation therapy; Diagnosis-therapy integration.

## 1. Introduction

Originated from somatic mutations, the tumor cells generally proliferate and differentiate continuously under the regulation of various growth factors as well as the oxygen and nutrients provided throughout the blood vessels [1-4]. At the early stage when the death and the regeneration of tumor cells are nearly balanced, the existing vascular network near the tumor tissues supplies nutrients and carries away metabolites. Afterwards, a new tumor vascular network generates around the tumor tissues due to the vascular endothelial growth factor secreted by the tumor cells [5]. Then, the rapidly increasing tumor cells would diffuse to other parts via the tumor-induced angiogenesis. Clearly, the blood vessels play a significant role on the growth and metastasis of the tumor [6].

Considering the dependency of the tumor growth on the blood vessels, the treatment of vascular targeting therapy for starving the tumors to death has increasingly attracted the attentions of researchers across the world [7, 8]. One of such strategies is to prevent the tumor-induced angiogenesis by the angiogenesis inhibitors, which was originally proposed by Folkman in the early 1970s [9]. Ever since then, various inhibitive drugs have been developed [10-13], and such embolization therapy has also been combined with chemotherapy or radiotherapy for better treatment [14]. However, the angiogenesis inhibitors have to be used for a long time. And some studies found that within this period, the tumor itself would gradually generate resistance to these drugs, which makes it not as effective as expected [15]. As another approach focusing on blocking the existing vessels, colchicine was reported to destroy the vessels, yet with rather high toxicity [16]. Afterwards, some other vascular disrupting agents were studied including small molecular agents [17, 18].

Except for these drugs, there are some other physical embolization ways to offer the possibility of starving the tissues or tumor to death by means of embolic agents, such as autologous clots, coils, gelfoam, balloons, glue, nanoparticles, microspheres embolic agents and so on [19-23]. However, the failure to embolize the vessels may often occur. One key reason lies in that some solid embolic agents cannot conform well to the vessels, leaving certain space at the interface for nutrients to leak in. Another existing issue is that some embolic agents are easily rushed away by the blood flow, such as the balloons. What's more, the catheters or other more complex equipments are needed in order to place some of embolic agents into the target positions. Different from these approaches, gas embolotherapy uses gas bubbles to occlude the blood flow [24]. The bubbles originate as small diameter liquid droplets of dodecafluoropentane (DDFP) mixed in saline and albumin. After injected into the vascular system, these droplets are small enough to pass through the capillary beds. At the strategic location, the droplets are vaporized into larger bubbles through acoustic wave to occlude flow. However, the entry of gas into the vessel system is a risk in virtually all areas of clinical care [25].

In view of the above tumor treatment situations, this paper is dedicated to proposing and demonstrating an alternative vascular embolization therapy, the body temperature liquid metal, as a kind of unconventional vessel embolic agent for starving the target tissues or tumors to death. Here, to emphasize the method itself, although plenty of liquid metal candidates can serve as the treatment agents, we mainly tested for the first trial the pure gallium and $Ga_{75.5}In_{24.5}$ alloy as the illustration examples. To characterize the physical behaviors of such metal agent, the DSC experiments on the liquid metal gallium were conducted, which revealed the thermal properties of the gallium and verified its feasibility for the delivery of therapy. The cytotoxicity of the gallium and indium *in vitro* were also evaluated with the tests of CCK-8 and flow cytometry. As an additional advantage, the liquid metal in vessels showed high image contrast under the X-ray scan, which indicated its great potential of the diagnosis-treatment integration. For further investigation, we experimented on several different tissues or transplanted tumor to demonstrate the liquid metal embolization effect in vivo. Finally, we presented the theoretical modeling and simulation of tumor growth affected by liquid metal enabled blockage of blood vessels near the tumor. This study opened the way of applying the liquid metal vascular embolization for a potential tumor starvation therapy in the coming time.

## 2. Principle and methods

## 2.1 Basic feature of liquid metal vascular embolic therapy

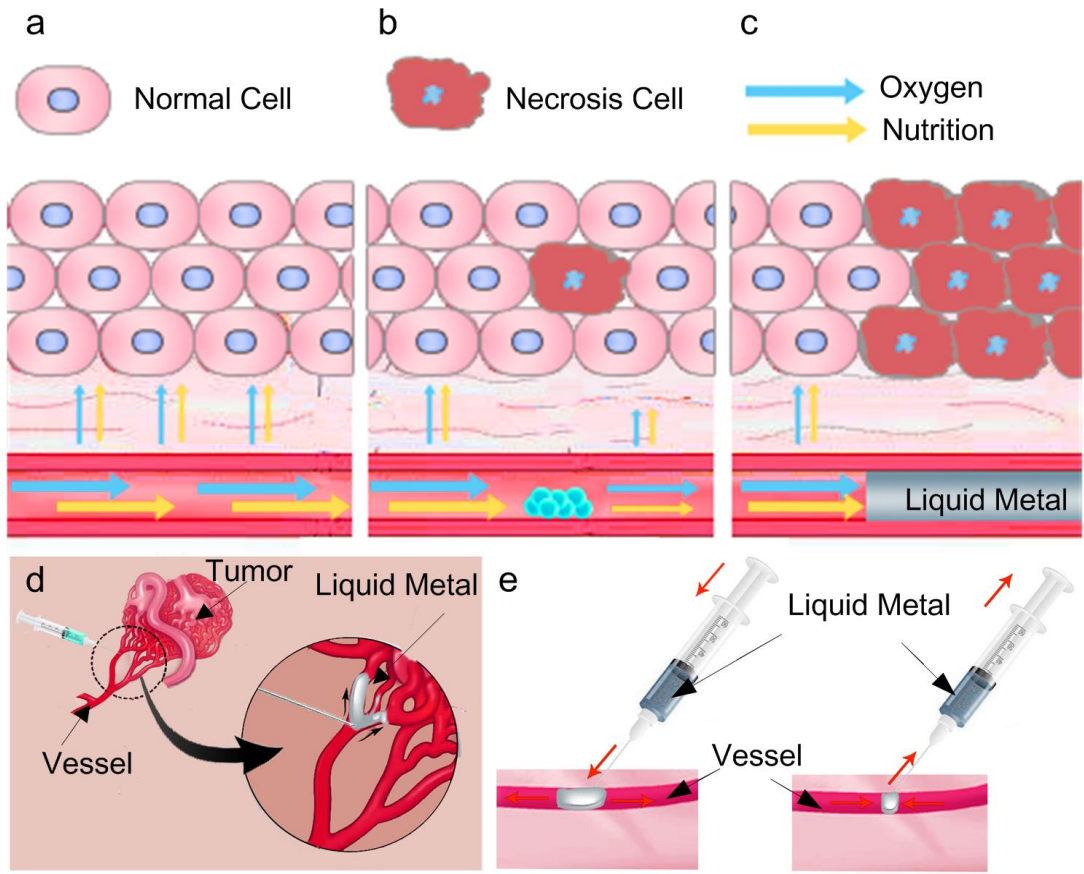

**Figure 1** The principle illustration of liquid metal based tumor vascular embolization therapy. **(a)** The oxygen and nutrition supplies from the vessels to tissues without any embolic agents. **(b)** The incomplete occlusion. **(c)** The complete occlusion by the liquid metal. **(d)** The physical occlusion of the blood supply to tumor. **(e)** The liquid metal could be injected into or sucked out of the vessel in case of need.

As a soft, compliant and fluidic material, the liquid metal can be easily injected and filled into the vessels and capillaries sufficiently, which stops the blood flow and all the related physiological events thus enabled because of its high density and fluidity. Fig. 1 illustrates the basic principle of the liquid metal embolization therapy and surgical operation of its injection and removal. As is widely known, tissues acquire oxygen and nutrition from the nearby feeding vessels (Fig. 1a). Though the vascular blockage can shut down the blood supply, yet the stream is still able to pass through the vessels interior when the embolic agent fails to completely occlude the vessels. As a result, the reduction of blood flow does not guarantee a strong enough effect on starving the necrotic cells (Fig. 1b) and the embolic agent can even possibly be rushed away within a short period after it was injected. Being able to maintain in liquid state at body temperature, the liquid metal agent can easily shape itself along the vessels and fully occlude the channels due to its perfect compliance (Fig. 1c). Compared with the normal tissues, the tumor is more dependent on the nutrients supplies from the surrounding vessels. After injected into the vessels, the liquid metal results in a physical occlusion and thus leads to tumor regressions (Fig. 1d). Another merit of the current therapy still lies in that, theoretically, the liquid metal could also be well sucked out in case of need after completing its treatment role (Fig. 1e).

*2.2 Characterization of the physical properties of the liquid metal agents*

We measured the thermal property of gallium (99.999%, purchased from Anhui Rare New Materials Co. LTD, China) with the differential scanning calorimetry device (DSC 200 F3 Maia, Netzsch Scientific Instrument Trading Co. Ltd, Germany). The 14.63mg of gallium was put into the alumina crucible with another empty one as the reference, which were both placed in a heating furnace. The test temperature range included three cycles, which were set as 40 ℃ → 60 ℃ for 1min → -60 ℃ for 1min → 60 ℃ for 1min → -60 ℃ for 1min → 60 ℃ for 1min → -60 ℃ for 1min → 60 ℃ for 1min → 45 ℃ → 70 ℃, and the heating and cooling rates were both prescribed at 10 ℃ /min. The results were analyzed by the software of NETZSCH Proteus Thermal Analysis.

*2.3 Evaluation of the metal agents cytotoxicity in vitro*

Two tests including the cell counting kit-8 (CCK-8) and the flow cytometry were applied on the mouse embryonic fibroblast cells (NIH3T3) to evaluate the cytotoxicity of the liquid metal.

A drop of the gallium was put on the six-well plates filled with culture solution, soaking for 24 hours and 48 hours in the incubator. Meanwhile, a piece of polished and sterilized indium and copper were soaked under the same condition.

CCK-8 method: Cells at logarithmic growth phase were treated by trypsin and tuned into single cell suspension with a concentration of 50000/ml, then seeded in a 96-well plate with 100μL in each well. After being cultured for 24 hours, the cells were observed under the microscope and the samples in bad state were eliminated. The original culture solution was removed and the wells were washed with PBS solution. Then, the soaked solution of gallium, indium, copper and the original culture solution (control group) were added into the wells with 5 parallel samples in each group. After being cultured for 24 hours, each well was washed and added with a mixture of 100μL fresh culture solution and 10μL CCK-8. The 96-well plate were put into the incubator for 2 hours, then optical density (OD) at 450nm wavelength were measured with multifunctional microplate reader (EnVision, USA PerkinElmer). The cell viability was calculated according the following equation:

$$\text{Cell viability} = (OD-OD_0)/(OD_c-OD_0)) \qquad (1)$$

where, $OD_c$ was the OD value of the control group and $OD_0$ was the OD value of black wells without any cells.

Flow cytometry method: The cells were seeded into the sterile dish of 60mm diameter with 300000 cells each. After cultured for 24 hours in the incubator, the original culture medium was removed and the wells were washed with PBS solution. Then, the three soaked solutions and the original culture solution were added separately. After cultured for another 24 hours, the cells in each dish were collected into the centrifuge tubes. By centrifuge at 1500r/min for 5min, the supernatant was removed and 1mL cell staining buffer and 5uL PI staining solution were added, then a 30-min ice bath followed. Finally, the cell solution were placed in the flow cytometer, detecting the red fluorescence by flow cytometry (BD Calibur, the USA BD).

## 2.4 Liquid metal angiography of the tumor vessels

It was recently found that, different from other embolic agents, liquid metal has excellent visibility under the X-ray [26]. Thus in a further study, the tumor vessels injected with the liquid metal went through a CT scan to investigate the distribution of the material. In order to assess this special characteristic, several representative objects such as *in vitro* swine kidney, a healthy 8-week-old female CD1 mice and an 8-week-old female Barb/c nude mice (purchased by the Center of Biomedical Analysis of Tsinghua University) with a tumor were selected to serve as the experimental testing cases.

About 0.8mL liquid metal gallium was infused into the artery branches of the swine kidney. Then the filled kidney was X-ray photographed by a micro-CT scanner (XM-Tracer-130, from Institute of High Energy Physics, Chinese Academy of Sciences).

The mouse has both arteries broad enough for liquid metal $Ga_{75.5}In_{24.5}$ alloy infusion and thin capillaries to display the alloy's high contrast. First, the mouse was excessively anesthetized to death, then the heart was exposed by cutting open its chest. A thin syringe needle was used to penetrate into the left ventricle of the mouse heart and it was clamped to prevent leakage and reverse flow. With a slight cut near the right atrium, the liquid metal alloy could finally be infused into the circulatory tunnels and flow throughout the body. When the silvery alloy flows to the right heart, it was supposed to finish the infusion and that the sample had been stuffed with the contrast agent. Then the sample was taken to the imaging instrument (Super Nova CT, from Technical Institute of Physics and Chemistry, Chinese Academy of Sciences). Images with higher resolution were obtained from the micro-CT (XM-Tracer-130, from Institute of High Energy Physics, Chinese Academy of Sciences).

The structures of the vessels in the tumor tissues and normal tissues may be different. In order to offer more evidences on various tissue vasculatures, we also managed to infuse the liquid metal $Ga_{75.5}In_{24.5}$ alloy into the tumor vessels. The mouse breast cancer cells (EMT6, $1 \times 10^6$ cells) were injected directly into the subcutaneous at the back of the nude mice. The tumor reached approximately a diameter of 5mm in two weeks and the vessels were visible on the skin. After the mice was anesthetized with intraperitoneal injection of 0.2mL 1% (g/L) sodium pentobarbital (Nembutal), the liquid metal was injected directly into a tumor vessel via a special syringe needle with an inner diameter of 0.3mm. Afterwards, the tumor region went under a CT scan (Super Nova CT, from Technical Institute of Physics and Chemistry, Chinese Academy of Sciences) for visualizing the vessels injected with liquid metal.

*2.5 Liquid metal as embolic agents of blood vessels in vivo*

A two-month-old male New Zealand rabbit (purchased from the Center of Biomedical Analysis of Tsinghua University), weighted about 2.2kg, were used as the experimental subject, because the vessels in the ears are broad and visible. Anesthetized with intraperitoneal injection of 4% (g/L) sodium pentobarbital (Nembutal) by the standard of 1 ml/kg, the rabbit was injected with liquid metal gallium into the vein of one ear. It was observed that the material filled the vessels at the rim and the tip of the ear quickly. The other ear was used for contrast. The infrared camera (Thermovision A40, FLIR, Wilsonville, OR) was adopted to record the temperature changes of the rabbit ear during the operational process. Afterwards, the rabbit was observed for half an hour and then put back to the caring center. The rabbit continued to be observed every day. Blood analysis were carried out before the experiments and 1 day, 3 days, 7 days afterwards with the automatic hematology analyzer (BM800), each time with 3 repeated trials. After the occurrence of necrosis at the ear tip, the rabbit was sacrificed through standard procedure. Then the tissues of both the necrotic region and the same position of the other ear were surgically cut and immediately restored in formalin for 24 hours. The samples were then sectioned, stained with HE methods (Hematoxylin-eosin) and observed under the microscope.

Afterwards, we injected the liquid metal gallium into the vessels of the mouse tail and the rabbit leg in the same way. All experiments above have been approved by the Ethics Committee of Tsinghua University, Beijing, China under contract [SYXK (Jing) 2009-0022].

## *2.6 Theoretical modeling and simulation of tumor growth with liquid metal embolization*

Although the liquid metal embolization of blood vessels *in vivo* was implemented on the vessels of rabbit ears and legs, a case of tumor might be more illustrative. Here, a theoretical modeling and simulation of tumor growth affected by the metal blockage of blood vessels nearby was (see Fig. 2) carried out to interpret the starvation effect of the current therapy.

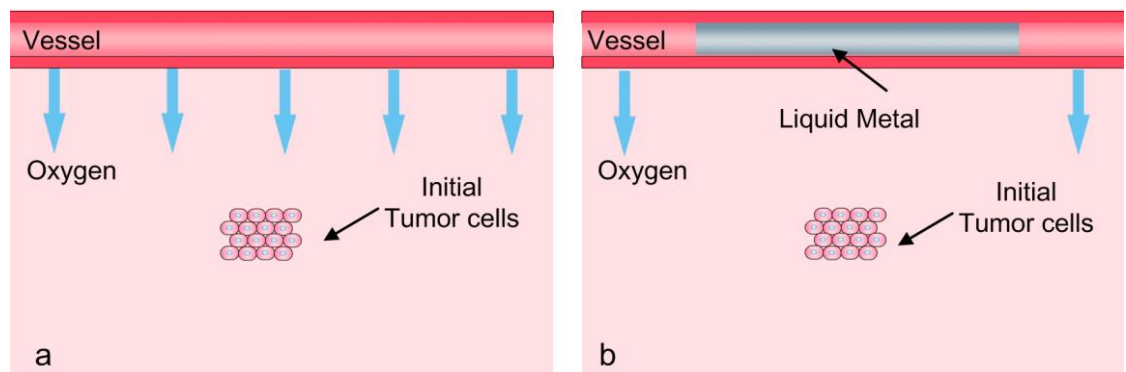

**Figure 2** Illustration of the theoretical tumor growth model. **(a)** Without the liquid metal in the vessel. **(b)** With the liquid metal embolic agent in the vessel.

Without losing any generality, there are generally four dependent variables to dominate the tumor growth model, including cancer cell density, oxygen density, extracellular matrix density and matrix degrading enzymes. The functions of each variable (further details about the functions could be seen in Alexander's work [27]) in non-dimensional are expressed as

$$\frac{\partial n}{\partial t} = d_n \nabla^2 n - \rho \nabla \cdot (n \nabla f) \quad (2)$$

$$\frac{\partial o}{\partial t} = d_o \nabla^2 o + \nu f - \omega n - \varphi o \quad (3)$$

$$\frac{\partial f}{\partial t} = -\eta m f \quad (4)$$

$$\frac{\partial m}{\partial t} = d_m \nabla^2 m + \kappa n - \sigma m \quad (5)$$

where $n$, $o$, $f$ denote the density of cells, oxygen and extracellular matrix respectively, and $m$ denotes the matrix degrading enzymes.

The oxygen is supplied by the vessel and we make the assumption that the oxygen concentration is zero at the metal blockage position. The probabilities of movement of an individual cell are generated by discretizing the Equation (2).

**Table 1** The non-dimensional parameter values used in the simulations

| Parameters | Value | Parameters | Value |
|---|---|---|---|
| $d_n$ | 0.0005 | $\kappa$ | 1 |
| $d_o$ (out of /in the tumors) | 0.05/0.025 | $\sigma$ | 0 |
| $d_m$ | 0.0005 | $\nu$ | 0.5 |
| $\rho$ | 0.01 | $\omega$ | 0.57 |
| $\eta$ | 50 | $\varphi$ | 0.025 |

Besides, it is assumed that the cancer cells migrate through a combination of diffusion and haptotaxis as well as undergoing proliferation. And each cell is allowed to migrate when the number of neighbor cells is more than one. Cell death can occur due to lack of oxygen. In the model, we also make the assumption that each cell produces two daughter cells when the cell has reached maturity. If no empty space around the parent cell exists, the cell becomes quiescent until space appears. We assume that quiescent cells only consume half amount of oxygen of the active tumor cells. The tumor cells would not only maintain one phenotype, but also becomes more unstable and more aggressive phenotypes due to mutations. In the model, we mainly consider three phenotypes. The irreversible mutations occur at the mitosis stage.

The computation domain is divided by 200×200 grids with a space step of $h = 0.005$ and a time step of $k = 0.0001$. Initially, 36 tumor cells are in the center. The initial value of oxygen density, extracellular matrix density and matrix degrading enzymes are 1, 1 and 0, respectively. The non-dimensional parameter values used in the simulations are listed in Table 1 [27].

## 3. Results

### *3.1 Physical properties of the liquid metal*

Fig. 3 depicts the DSC curves of the gallium in reciprocally reverse processes. In either the cooling or the heating process, the measurement goes for three trials and the results are almost identical. There is a very sharp exothermic peak over the cooling process, and the freezing temperature is 4.5 ℃, while the endothermic peak is also steep in the heating process and the melting point is 28.3 ℃. The curves indicate an easy transition between the solid and liquid phase of the gallium, which is an excellent property for the phase control. What's more, owing to its sub-cooling effect, the material is able to maintain in the liquid state for a long time at the body temperature. Thanks to its liquid phase, compliance and fluidity, the material is able to fluently flow into and fill the tiny vessels including the capillaries whose diameter is at a scale of micrometers.

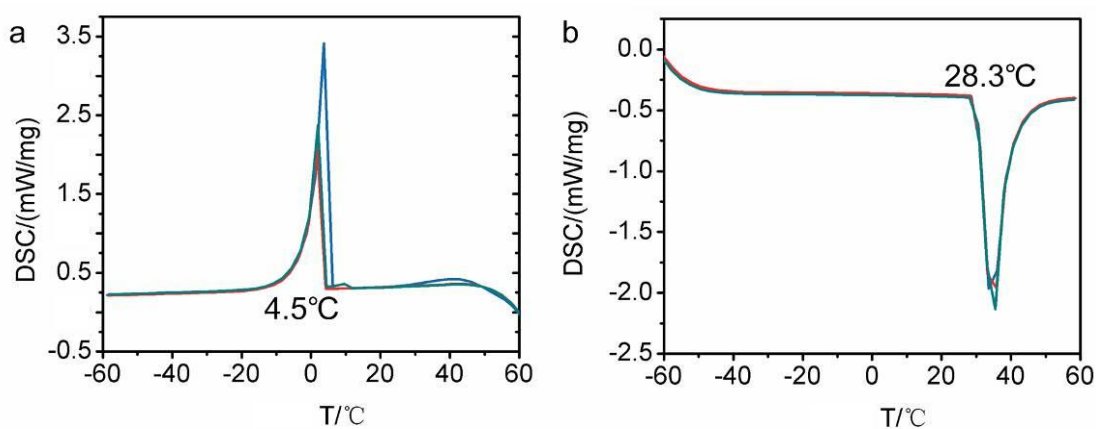

**Figure 3**　The DSC curves of the gallium. **(a)** Cooling process. **(b)** Heating process.

*3.2 Cytotoxicity of the liquid metal in vitro*

Fig. 4a shows the viability of the cells after cultured in different metal soaked solutions with CCK-8 method. The cell viability in the copper soaked solution in 24 hours was measured as 7%, and almost all died in 48 hours. The cell viability of the gallium soaked solution in 24 hours and 48 hours were 100.6% and 75.7%, and that of indium were 107% and 79.7%, respectively. According to the cytotoxicity evaluation criteria of the national standard, cell viability above 75% is qualified and safe. The viability of the cells can also be visually demonstrated through the microscope, which is shown in Fig. 4b. The cells of the gallium, indium soaked solution and original culture solution were exactly of the right size and shape large, while those in the copper soaked solution distributed sparsely and the shapes were relatively round, which means most of them were dead.

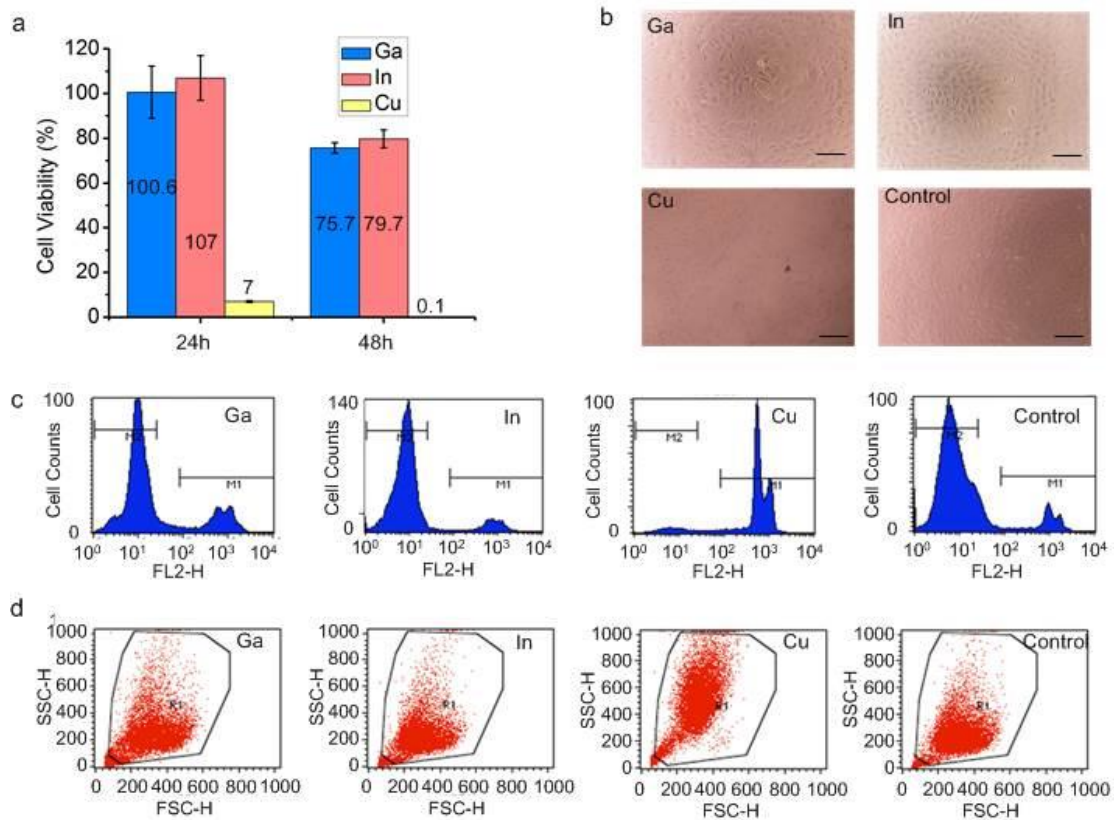

**Figure 4** Cytotoxicity in vitro. **(a)** The cell viability after cultured in different metal soaking solution with CCK-8 method. **(b)** The cellular morphology in each experimental group observed under the microscope. **(c)** The cell apoptosis with flow cytometry method. **(d)** The scatter images with flow cytometry method.

Fig. 4c indicates cell apoptosis of each experimental group. In the PI fluorescence intensity histograms, M1 stands for apoptosis cells while M2 represents living cells. Compared with the control group, the relative apoptosis rates of gallium and indium group was 10.18% and 7.9% respectively while 78.44% for the copper group. In the scatter images in Fig. 4d, it can be further found that the strength of forward scattered light reduced and the one side scattered light increased in the copper soaked solution, which also indicated that there were more apoptosis and necrosis cells. However, the gallium and indium group showed tiny difference with the control group.

These experiments have revealed that the cytotoxicity of the gallium and indium ion are relatively low and their applications on localized severe disease treatment are clinically acceptable, given its unique values as will be revealed in later sections.

## 3.3 Liquid metal angiography of the tumor vessels

Endowed with a high density, the liquid metal is an excellent contrast agent for the vessel-like tissues under the X-ray [26], which is a unique advantage compared with most other embolic agents.

The X-ray images of renal artery of the *in vitro* swine kidney are presented in Fig. 5a. Clearly, a whole branch of the renal artery network is intact and the texture of the small vessels is rather clear. As shown in Fig. 5b, the vessels are separated with the other tissues obviously in highlight in the CT scan reconstruction of a whole-body infused mouse. In the image, it is easy to identify the liquid metal's flow path throughout the carotid, limb and tail vessels even to the end tips. Besides, the branches in the lung, head and abdominal organs can also be distinguished with ultra-high contrast and clarity. In order to offer more evidence of the image guidance ability of the liquid metal alloy enhancement, Fig. 5c provides the closer X-ray photographs of vascular network in the abdomen of the mouse which has plenty of even tinier vessels. In the figure, the sensor focused in a square field with each side 6.6mm long, and the resolution is thus set as 13μm/pixel. Accordingly, it can be referred to that the thinnest visible vessel is less than 30μm wide, which means a tremendous improvement of the blood vessel visualization. Besides, the tumor vessels filled with liquid metal are highly visible under the CT scan as expected. As can be seen, the tumarized vasculature appeares somewhat irregular (Fig. 5d) as compared with normal tissue vessels (Fig.5a). This clearly reflected the growth behavior of such diseased vessels.

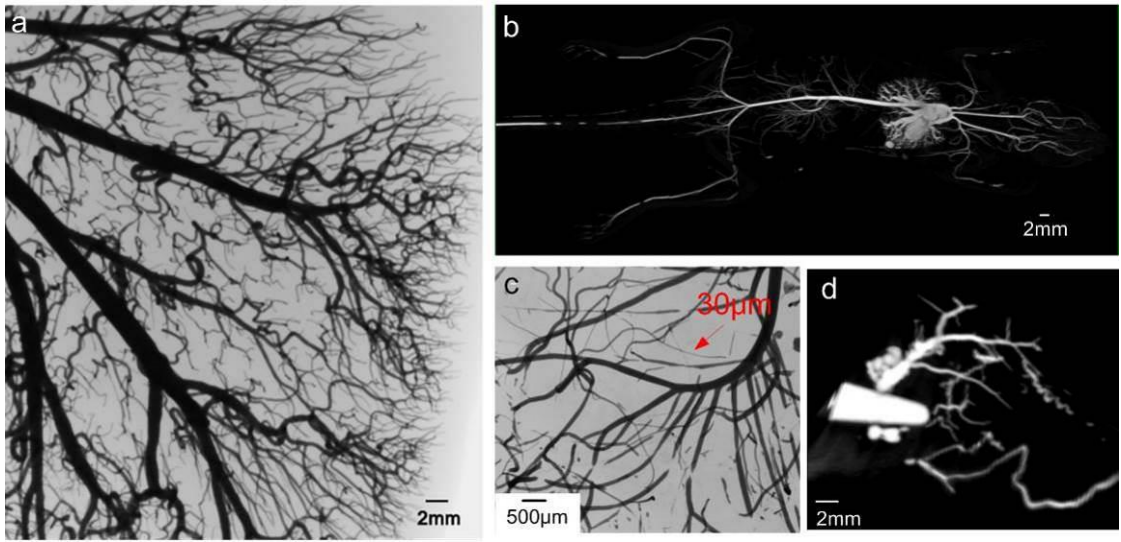

**Figure 5** The images of vessels filled with the liquid metal under the X-ray. **(a)** The X-ray image of renal artery of the *in vitro* swine kidney. **(b)** The top view of a liquid metal infused mouse in a whole-body CT scan and 3D reconstruction. **(c)** The X-ray image of vessels in the mouse abdomen with higher resolution. **(d)** 3D reconstruction image of tumor vessels.

*3.4 Liquid metal embolization to starve target living tissues to death*

As shown in Fig. 6a, the liquid metal has filled the rabbit ear vein at the rim and the tip. Without abnormal reactions, the rabbit showed obvious symptoms of necrosis only in the ear tip in three weeks, which finally come out like a dry leaf. The necrotic changes appeared not so serious at the rim of the ear, which might be due to the supply of certain surrounding tiny vessels. However, the very side of the ear had been crimped, which was a reflection of the liquid metal obstruction.

As an attendant phenomenon, the temperature of the liquid metal obstructed region would decrease due to stop of the blood flow. As a special evaluation of the material's influence, the temperature distributions in 0h, 24h, 48h and 72h from the thermal infrared images are depicted in Fig. 6b. Clearly, at the moment of the injection, the blood was immediately blocked, causing the temperature rise in the upstream and decrease in the downstream regions. With the time went on, the area of the low temperature region shrunk gradually, which indicated the concentration of the material to the tip.

In order to provide a biological inspection, Fig. 6c displays the sections of both the necrotic and normal tissues. It can be seen that near the liquid metal blocked vessels, the surrounding area tends to be emptier and the cells are much smaller than the situation in the similar region in a normal tissue. This shows the damage effect of the liquid metal embolization treatment.

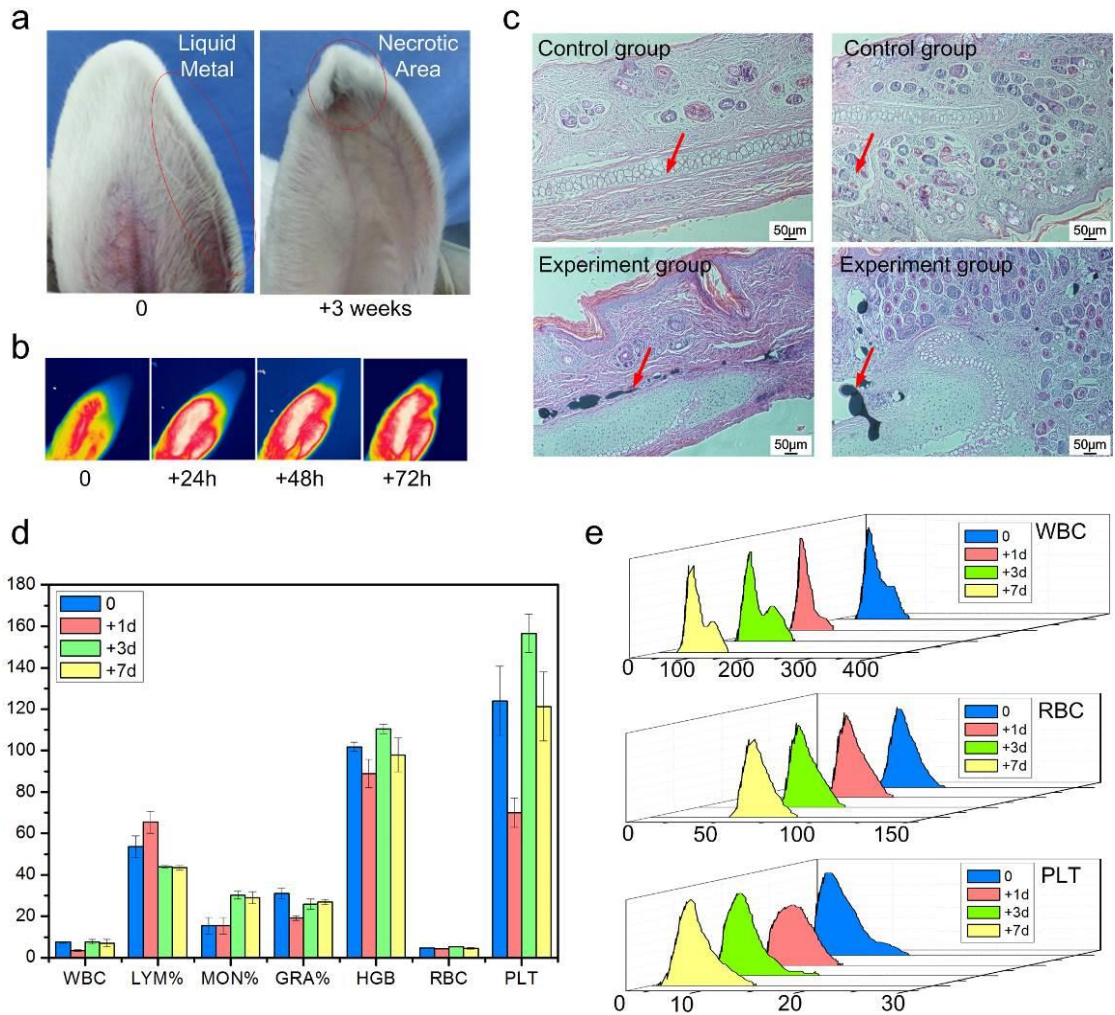

**Figure 6** The effects of liquid metal embolization at the rabbit ears. **(a)** The symptoms of necrosis in the ear. **(b)** The temperature distribution changes after 0h, 24h, 48h and 72h. **(c)** Comparison between the necrosis tissue sections and control normal tissue section. **(d)** The results of WBC, LYM%, MON%, GRA%, HGB, RBC and PLT* changed over time. **(e)** WBC, RBC and PLT dynamic histograms distribution.

*WBC: white blood cell; LYM: lymphocyte; MON: monocyte; GRA: granulocyte; HGB: hemoglobin; RBC: red blood cell; PLT: platelet.

Other than the morphologic analysis, blood tests were also carried out in a week and seven parameters including WBC, LYM%, MON%, GPA%, HGB, RBC and PLT are shown in Fig. 6d. Fig. 6e shows the WBC, RBC and PLT dynamic histograms distribution, respectively. As can be seen from the figure, the second peak of WBC distribution was lowered after one day but returned to normal after 3 days, which indicated the reduction of the neutrophil granulocytes. These blood analysis results demonstrated that the blood parameters would soon return to normal level after the liquid metal gallium was injected into the blood.

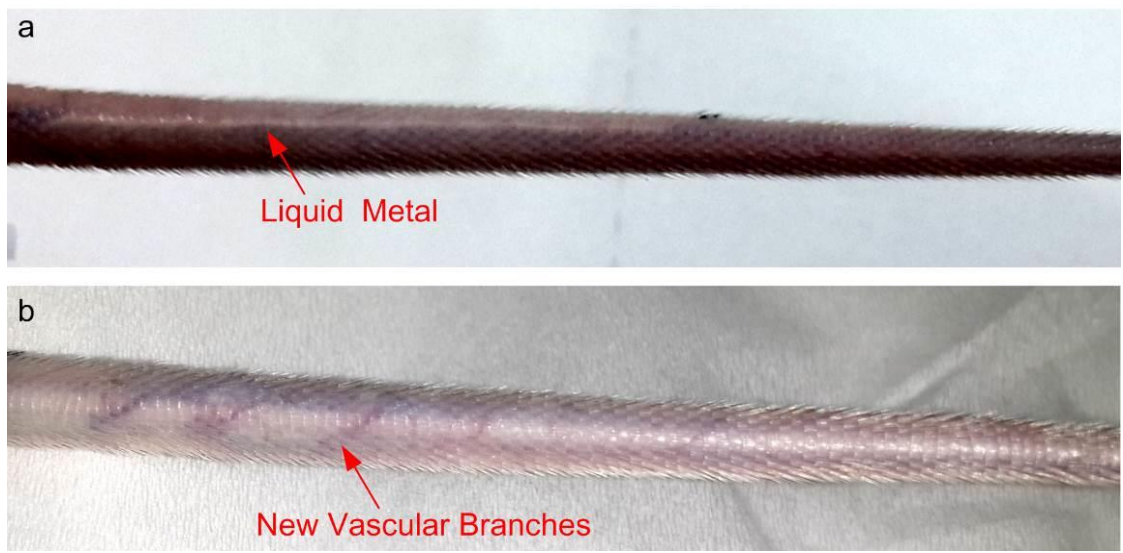

**Figure 7** The effects of liquid metal embolization at the mice tail. **(a)** The liquid metal injected into the tail vein. **(b)** The regenerated vascular branches.

It should be pointed out that, not all the injections will be able to completely fill the vessels and thus lead to damage of the target tissues. The output in fact depends on the specific surgical injection and delivery of liquid metal. To explain this point, we select the tail vessels of the mice as an additional test object to evaluate the blockage effect of liquid metal embolization on the tissues in vivo. As shown in Fig. 7a, three tail vessels of the mice were filled with liquid metal gallium. However, there was no visible necrosis after three weeks later. Moreover, certain new small blood vessels were even observed to regenerate (Fig.7b). The reason can be interpreted as follows. With a single flow direction and different from former cases, it was more difficult in current situation to fully block the femoral vessels at the tail of the mice. This led some of the liquid metal to flow away and no visible necrosis occurs at the target tissues over the whole experiments. Besides, these regenerated blood vessels near the tail vessels have further contributed for the nutrition and oxygen supply. This reminds us that, without appropriate administration of injecting the liquid metal agent, the embolization effect as anticipated before may not always necessarily be guaranteed. Therefore, for therapeutic purpose, careful treatment planning should be made in advance to surgically deliver liquid metal into the target vessels along specific directions. Lastly, it was also worth mentioning that these mice filled with liquid metal were kept alive for a rather long time like several weeks. Such findings indicate the safety and non-toxicity of the liquid metal materials staying inside the vessels on condition that they will not cause physically physiological danger.

## *3.5 Theoretical results of tumor growth with liquid metal embolic agents*

Theoretical simulations on the two cases with and without liquid metal embolic agents in the vessels near the tumor have been implemented to interpret and evaluate the performance of tumor growth suppression by the present method. Fig. 8 depicts the changes of oxygen concentration near the tumor at six times (5 days, 9days, 11days, 13days, 15days and 17days later). The tumor growth consumes oxygen and the oxygen concentration in tumor center is the lowest. In order to maintain growth, the tumor needs adequate oxygen supply by the vessel. However, as a result of liquid metal blockage in the vessel, the oxygen concentrations of the nearby tissues are decreasing. Fig. 9 shows the predicted results for tumor growth. Different color represents different phenotype cells, and red is the most aggressive cells. The tumor grows towards the vessel since there exists only one single vessel on one side. The tumor cell distribution shows a dead central region, which is due to lack of the oxygen as shown in Fig. 8. For the case without liquid metal embolic agent, the dead central region occures after the 17 days, while for the case with liquid metal embolic agent, the dead central region occurs after the 11 days. As the time goes on, the dead central regions continue to enlarge.

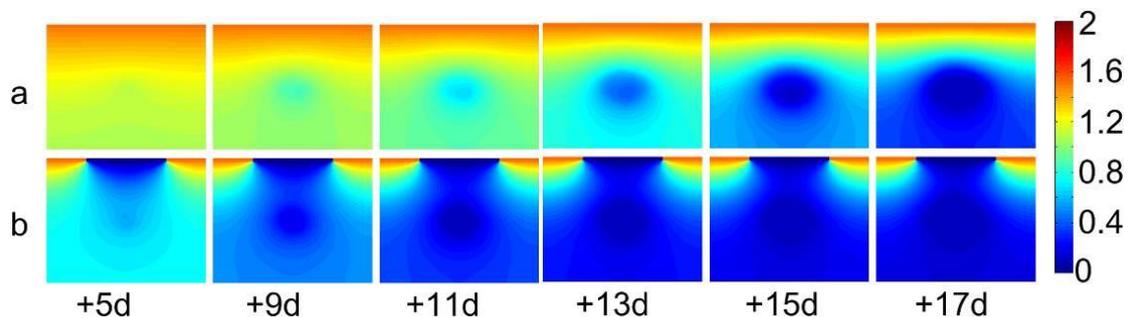

**Figure 8** The oxygen distribution with the change of time. **(a)** Without the liquid metal in the vessel. **(b)** With the liquid metal embolic agent in the vessel.

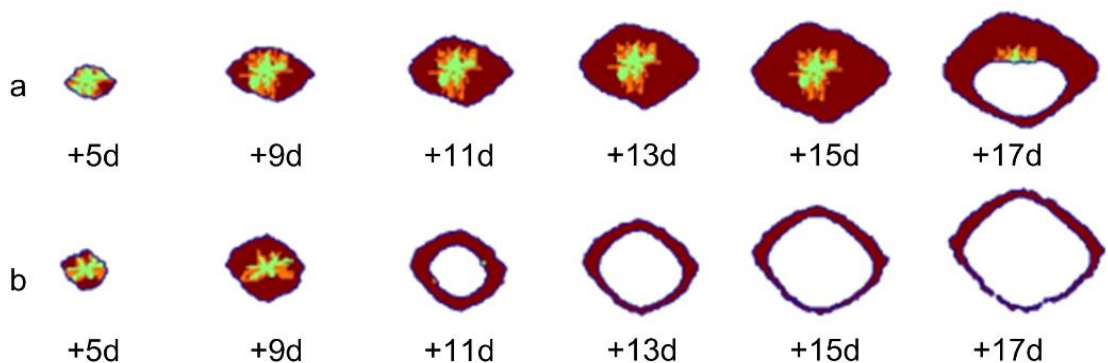

**Figure 9** Spatial distribution of tumor cells with the time. **(a)** Without the liquid metal in the vessel. **(b)** With the liquid metal embolic agent in the vessel.

## 4. Discussion

As is well known, the abundant vessels in the tumor tissues make it difficult for tumor treatment. Curatie surgery can easily lead to a massive hemorrhage, and large blood vessels would produce localized cooling in heated tissues during tumor hyperthermia, for instance. In this study, body temperature liquid metal yields a promising injectable tumor treatment. With its easy phase conversion, such material owns the capability to flow into tiny vessels at liquid state and stay at the target tissues at the solid state. Once the liquid metal was injected into the vessels, the nearby tissues would be possibly starved to death due to cutting off the feeding route of the nutrition and oxygen supplies.

Though the unprecedented application of the liquid metal on the tumor treatment seems to be beyond imagination, there remains a wide range for the further utilization of the material. In the present experiments, only gallium and $Ga_{75.5}In_{24.5}$ alloy were tested for brief. In fact, the melting points of the liquid metal could be modified by adding some other metals into alloys and adjusting the compounds ratio.

As a metal, the gallium is endowed with good electrical and thermal conductivity. Hence on the base of the blocking therapy, newer treatment might still be developed through combining the electrical and thermal stimulation together. We could even add certain substances into the liquid metal to obtain or enhance specific properties. Ma and Liu proposed a method to fabricate liquid metal with desired properties by loading with nanoparticles [28]. These nanoparticles include magnetic oxides such as $Fe_3O_4$, metallic particles and some semi-conductive particles, which further expands the application of liquid metal as the embolic agents. What's more, it is possible to combine this approach with chemotherapy or radiotherapy by injecting mixed liquid metal with chemotherapy and radiotherapy substances.

In this study, it is worth emphasizing the role of the liquid metal angiography, which reveals an extra advantage of the vessel blocking therapy. The visibility of the material under the X-ray has provided the possibility for monitoring the flow process along the gallium injection and evaluating vascular changes and damages. The combination of the liquid metal blocking and the X-ray imaging even opens a promising approach to the investigation of the tumor vessel growth and distribution, which also indicates a great potential of the diagnosis-treatment integration.

However, there still exist many issues to be solved in order to further improve the efficiency of the liquid metal embolization. Considering the difficulties to inject the liquid metal into the artery due to fast blood flow there and deep position in the tissues, we almost administrated these experiments on the veins. However, as we all know, the blood in artery flows into the capillary vessel, while that in vein blood flows out of the capillary vessel. Therefore, it is better to inject the liquid metal into the artery to occlude the vessel. Besides, the regeneration of the tumor vessel network under the obstructive condition is also a common challenge in the blocking therapy.

The safety is also a non-negligible issue in this new conceptual blood vessel embolization strategy. As mentioned above, some of the liquid metal could be washed away from the target tissues since injection from the veins. As is shown in Fig. 10, the liquid metal has entered the heart and lungs of the rabbit in the experiments of leg infusion yet the rabbit is still alive. Therefore it is very important to adopt measures to prevent this problem from happening in further studies. One possible solution is to apply proper pressure on the proximal end to slow the blood flow velocity. Another method to decrease the blood flow is to reduce the temperature of the blood in the target tissues, which could also speed up the solidification of the liquid metal. It also indicates that the *in vivo* application is highly sophisticated due to safety control and thus skilled surgical operation is required as well. In this paper, we only used gallium and $Ga_{75.5}In_{24.5}$ alloy as the embolic agents, whose melting point are both under the 37 ℃. Next, we will try to synthetize more other liquid metal alloys with a little higher melting point than 37 ℃ and test their embolization effects. In addition, a comprehensive evaluation of long-term toxicity still requests further research.

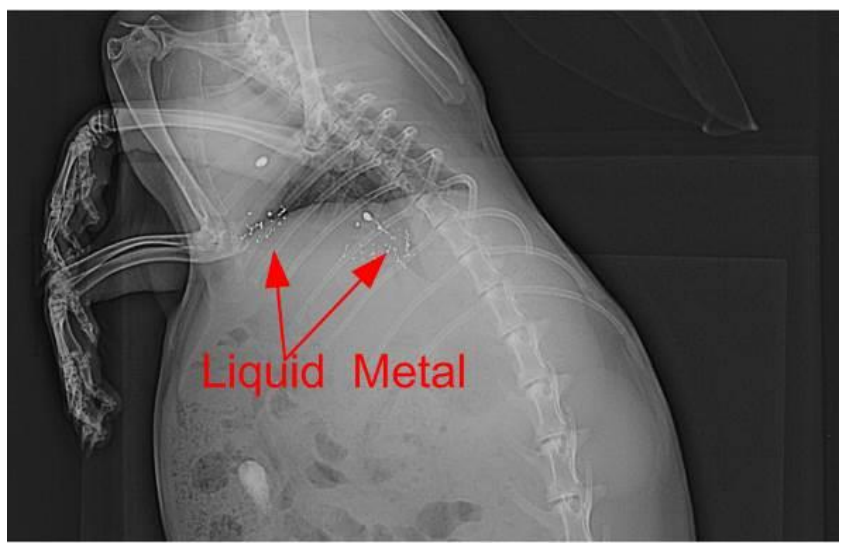

**Figure 10**  Traces of the liquid metal in the heart and lungs of the rabbit.

## 5. Conclusion

In summary, the liquid metal, as a fluidic material at body temperature, has been proposed for the first time as the blood vessel embolic agent to starve the necrotic tissues or target tumors to death. With its merits of easy phase transition and sub-cooling, this kind of material can be conveniently injected into the vessels with relatively easy operations. A series of *in vitro* experiments on the material cytotoxicity have been implemented to evaluate its safety and the results have indicated that both gallium and indium irons show low toxicity on normal cell growth. On considering its application against tumor or other severe diseases, such a performance is quite acceptable. As an extra advantage, the liquid metal is endowed with excellent contrast under the X-ray over the tissues, which provided a powerful soft tool for tumor vascular research and offered great potential to realize the diagnosis-treatment integration. Furthermore, both *in vivo* experiments and theoretical model simulations have preliminarily demonstrated the performance of the liquid metal to starve the tissues or tumors to death. Though still facing certain challenges, the unprecedented utilization of the liquid metal agent opens a new way for further practice in future tumor vessel blocking therapy.

## Acknowledgement

This work was partially supported by the NSFC under Grant 51376102.